\documentstyle[12pt]{article}
\begin{document}

\def\frak{\cal }
\def\Bbb{\bf }

\title{Forecast and event control\\On what is and what cannot be possible\\
Part I: Classical case}
\author{Karl Svozil\\
 {\small Institut f\"ur Theoretische Physik,}
  {\small Technische Universit\"at Wien }     \\
  {\small Wiedner Hauptstra\ss e 8-10/136,}
  {\small A-1040 Vienna, Austria   }            \\
  {\small e-mail: svozil@tuwien.ac.at}}
\date{ }
\maketitle

\begin{abstract}
Consequences  of the basic and most evident consistency
requirement---that measured events cannot happen and not happen at the same
time---are shortly reviewed.
Particular emphasis is given to event forecast and event control.
As a consequence, particular, very general bounds on the forecast and
control of events within the known laws of physics result.
These bounds are of a global, statistical
nature and need not affect singular events or groups of events.
\end{abstract}

\subsubsection*{Principle of self-consistency}
An irreducible, atomic physical phenomenon
manifests itself as a click of some detector.
There can either be a click or there can be no click.
This yes-no scheme is experimental physics in-a-nutshell
(at least according to a theoretician).
From that kind of elementary observation, all of our physical
evidence is accumulated.

Such irreversibly observed  events
(whatever the relevance or meaning of those terms are
\cite{wigner:mb,wheeler,greenberger2,hkwz})
are subject to the primary condition of {\em consistency} or
{\em self-consistency}:
{\em ``Any particular irreversibly observed event  either
happens or does not happen, but it cannot both happen and not happen.''}

Indeed, so trivial seems the requirement of consistency  that
David Hilbert polemicised against ``another
author'' with the following words \cite{hilbert-26}, ``...for me, the
opinion that the [[physical]] facts and events
themselves can be contradictory is a good example of thoughtlessness.''

Just as in mathematics, inconsistency, i.e., the coexistence of  truth
and falseness of  propositions, is a fatal property of any
physical theory. Nevertheless, in a certain very precise sense, quantum
mechanics incorporates inconsistencies in a very subtle way, which
assures overall consistency. For instance, a particle
wave function or quantum state is said to ``pass'' a double slit through both
slits at once, which is classically impossible. (Such considerations may, however, be
considered as mere trickery quantum talk, devoid of any operational meaning.)
Yet, neither a particle wave
function nor quantum states are directly associable with any sort of
irreversible observed event of physical reality.
We shall come back to a particular quantum case in the second part of this investigation.

And just as in mathematics and in formal logic it can be argued
that too strong capacities of intrinsic event forecast and intrinsic event control
renders the system overall inconsistent.
This fact may indeed be considered as one decisive feature in
finite deterministic (``algorithmic'') models \cite{svozil-93}.
It manifests itself already in the early stages of Cantorian set theory:
any claim that it is possible to enumerate the real numbers
yields, via the diagonalization method, to an outright contradiction.
The only consistent alternative is the acceptance
that no such capacity of enumeration exists.
G\"odel's incompleteness theorem   \cite{godel1} states that
any formal system rich enough to include arithmetic and elementary logic
could not be both consistent and complete.
Turing's theorem on the recursive unsolvability of the halting
problem  \cite{turing-36}, as well as Chaitin's $\Omega$ numbers
\cite{chaitin:92}
are formalizations of related limitations in formal logics, the computer sciences
and mathematics.

In what follows we shall proceed along very similar lines.
We shall first argue that any capacity of
total forecast or event
control---even in a totally deterministic environment---is
contradicting the (idealistic) idea that decisions between alternatives are possible;
stated differently: that there is free will.
Then we shall proceed with possibilities of forecast and event control
which are consistent both with free will and the known laws of physics.

It is also clear that some form of forecast and event control
is evidently possible---indeed, that is one of the main achievements of
contemporary natural science, and we make everyday use of it, say, by switching
on the light.
These  capacities derived from the standard natural sciences are
characterized by a high chance of reproducibility, and therefore do not
depend on single events.

In what follows, we shall concentrate on very general bounds of
capacities of forecast and event control;
bounds which are imposed upon them by the requirement of consistency.
These considerations should be fairly general
and do not depend on any particular physical model.
They are valid for all conceivable forms of physical theories; classical,
quantum and forthcoming alike.

\subsubsection*{Strong forecasting}
Let us consider forecasting the future first.
Even if physical phenomena occur
deterministically and can be accounted for ("computed") on a higher
level of abstraction, from within the system such a complete description
may not be of much practical, operational use
\cite{toffoli:79,svozil-unev}.

Indeed, suppose there exists free will. Suppose further that an agent could
predict {\em all} future events, without exceptions.
We shall call this the
{\em strong form of forecasting.}
In this case, the agent could freely decide to
counteract in such a way as to invalidate that prediction.
Hence, in order to avoid inconsistencies and  paradoxes,
either free will has to be abandoned,
or it has to be accepted that complete prediction is impossible.

Another possibility would be to consider strong forms of forecasting
which are, however, not utilized to alter the system.
Effectively, this results in the abandonment of free will,
amounting to an extrinsic, detached viewpoint.
After all, what is knowledge and what is it good for if it cannot be
applied and made to use?

It should be mentioned that the above argument is of an
ancient type \cite{martin}.
As has already been mentioned,
it has been formalized recently in set theory, formal
logic and recursive function theory, where it is called
``diagonalization method.''

In doing this, we are inspired by the recent advances in the foundations
of quantum (information) theory.
There, due to complementarity and the impossibility to clone
generic states, single events may have important meanings to
some observers, although they make no sense at all to other observers.
One example for this is quantum cryptography.
Many of these events are stochastic and are postulated to
satisfy all conceivable statistical laws (correlations are nonclassical,
though).
In such frameworks, high degrees of reproducibility cannot be
guaranteed, although single events may carry valuable information,
which can even be distilled and purified.

\subsubsection*{Strong event control}
A very similar argument holds for event control and the production of
``miracles'' \cite{frank}.
Suppose there exists free will.
Suppose further that an agent could entirely control the future.
We shall call this the {\em strong form of event control.}
Then this observer could freely decide to invalidate the laws of physics.
In order to avoid a paradox,  either  free will or some
physical laws would have to be abandoned, or it has to be accepted that
complete event control is impossible.

\subsubsection*{Weak forecast and event control}
From what has already been said, it should be clear that
it is reasonable to assume that {\em forecast and event control should be possible
only if this capacity cannot be associated with any paradox or contradiction.}

Thus the requirement of consistency of the phenomena seems to impose
rather stringent conditions on  forecasting and
event control. Similar ideas have  already been discussed in the context
of time paradoxes in relativity theory (cf. \cite{friedetal} and
\cite[p. 272]{nahin}, {\em ``The only solutions to the laws of physics
that can occur locally $\ldots$ are those which are globally
self-consistent''}).

There is, however, a possibility that the forecast and control of future
events {\em is} conceivable for {\em singular} events within the
statistical bounds. Such occurrences
may be ``singular miracles'' which are well accountable within
the known laws of physics. They will be called {\em weak forms of
forecasting and event control.}

It may be argued that, in order to obey overall
consistency, such a framework should not be extendable to any forms of strong forecast or
event control, because, as has been argued before, this could
either violate global consistency criteria
or would make necessary a revision of the known laws of physics.

The relevant laws of statistics (e.g., all recursively enumerable ones)
impose rather
lax constraints especially on finite sequences
and do not exclude local, singular, improbable events.
For example, a binary sequence such as
$11111111111111111111111111111111$
is just as probable as the sequences
$11100101110101000111000011010101$ and
$01010101010101010101010101010101$
and its occurrence in a test is equally likely, although  the
``meaning'' an observer could ascribe to it is rather different.
These sequences may be embedded in and be part of much longer stochastic
sequences. If short finite regular (or ``meaningful'') sequences are
padded into long irregular (``meaningless'') ones, those sequences
become statistically indistinguishable for all practical purposes
from the previous sequences.
Of course, the  ``meaning'' of any such sequence may vary with
different observers.
Some of them may be able to decipher a sequence, others may not be
capable of this capacity.

It is quite evident that per definition any
finite regularity in an otherwise stochastic environment should exclude
the type of high reproducability which one has gotten used to in the
natural sciences. Just on the contrary: single ``meaningful'' events
which are hardly reproducible might indicate  a new category of
phenomena which is dual to the usual ``lawful'' and highly predictable
ones.

Just as it is perfectly all right
to consider the statement ``This statement is true'' to be true, it may
be perfectly reasonable to speculate that certain events are
forecasted and controlled within the domain of statistical laws.
But in order to be  within the statistical laws, any such method
{\em needs not to be guaranteed} to work at all  times.

To put it pointedly: it may be perfectly reasonable to become rich,
say,
by singular forecasts of the stock and future values or in horse races, but
such an ability
must necessarily be not extendible, irreproducible and secretive; at
least to such an
extend that no guarantee of an overall strategy and regularity can be
derived from it.

The associated weak forms of forecasting and
event control are thus beyond any global statistical significance. Their
importance and meaning seems to lie mainly on a very subjective level of
singular events. This comes close to one aspect of what  Jung imagined
as the principle of ``synchronicity'' \cite{jung1},
and is dual to the more reproducible forms one is usually accustomed to.

\subsubsection*{Against the odds}
This final paragraph reviews a couple of experiments which suggest
themselves in the context of weak forecast and event control.
All are based on the observation whether or not an agent is capable
to forecast or control correctly
future events such as, say, the tossing of a fair coin.

In the first run of the experiment, no consequence is derived from the
agent's capacity despite the mere recording of the data.

The second run of the experiment is like the first run, but the {\em
meaning} of the forecasts or controlled events are different. They are
taken as outcomes of, say gambling, against other individuals (i) with
or (ii)
without similar capacities, or against (iii) an anonymous ``mechanic''
agent such as a casino or a stock exchange.

As a variant of this experiment, the partners or adversaries of the
agent are informed about the agent's intentions.

In the third run of experiments, the experimenter attempts to counteract
the agent's capacity. Let us assume the experimenter has total control
over the event. If the agent predicts or attempts to bring
about to happen a certain future event, the experimenter causes the
event not to happen and so on.

It might be interesting to record just how much the agent's capacity
is changed by the setup. Such an expectation might be defined from
a dichotomic observable
$$e(A,i) =\left\{
\begin{array}{cl}
+1&\qquad {\rm correct\; guess}\\
-1&\qquad {\rm incorrect\; guess}\\
\end{array}
\right.
$$
where $i$ stands for the $i$'th experiment and $A$ stands for the agent
$A$. An expectation function can then be defined as usual by the average
over $N$ experiments; i.e.,
$$E(A) = {1\over N}\sum_{i=1}^N e(A,i).$$

From the first to the second type of experiment it should become more
and
more unlikely that the agent operates correctly, since his performance
is leveled against other agents with more or less the same capacities.
The third type of experiment should produce a
total anticorrelation.
Formally, this should result in a decrease of $E$
when compared to the first round of experiment.

Another, rather subtle, deviation from the probabilistic
laws may be observed if {\em correlated} events are considered.
Just as in the case of quantum entanglement,
it may happen that individual components of correlated systems may behave
totally at random
exhibit more disorder than the system as a whole
\cite{nielsen-kem-2001}.

If once again one assumes two dichotomic observables $e(A,i),e(B,i)$
of a correlated subsystem, then the correlation function
$$C(A,B) = {1\over N}\sum_{i=1}^N e(A,i)e(B,i)$$
and the associated probabilities
may give rise to violations of the Boole-Bell inequalities---Boole's
{\em ``conditions of possible [[classical]] experience''}
\cite{Boole-62,Hailperin,pitowsky,Pit-94}
and may even exceed \cite{svozil-krenn}
the Tsirelson  bounds \cite{cirelson:80,cirelson:87,cirelson}
for ``conditions of possible [[quantum]] experience.''
There, the agent should concentrate on influencing the {\em coincidences}
of the event rather than the single individual events.
In such a case, the {\em individual} observables may behave perfectly random,
while the associated {\em correlations} might be nonclassical and even
stronger-than-quantum and might give rise to highly nonlocal phenomena.
As long as the individual events cannot be controlled,
this needs not even violate Einstein causality.
(But even then, consistent scenarios remain \cite{svozil-relrel}.)

In summary it can be stated that, although total forecasting and event control
are incompatible with free will, more subtle forms of these capacities remain
conceivable even beyond the present laws of physics; at least
as long as their effects upon the ``fabric of phenomena''
are consistent.
These capacities are characterized by singular events and not on the
reproducible patterns which are often encountered under the known laws
of physics.
Whether or not such capacities exist at all remains an open
question.
Nevertheless, despite the elusiveness of the phenomenology involved,
it appears not unreasonable that the hypothesis might be
testable, operationalizable and even put to use in certain contexts.


\end{document}